# Enhancing ferroelectric stability: Wide-range of adaptive control in epitaxial HfO$_2$/ZrO$_2$ superlattices

Jingxuan Li[1,†], Shiqing Deng[2,†], Liyang Ma[3,†], Yangyang Si[1], Chao Zhou[1], Kefan Wang[2], Sizhe Huang[1], Jiyuan Yang[3], Yunlong Tang[4], Yu-Chieh Ku[5], Chang-Yang Kuo[5,6], Yijie Li[1], Sujit Das[7], Shi Liu[3,*], Zuhuang Chen[1,*]

[1] State Key Laboratory of Precision Welding and Joining of Materials and Structures, School of Materials Science and Engineering, Harbin Institute of Technology, Shenzhen, 518055, China

[2] Beijing Advanced Innovation Center for Materials Genome Engineering, University of Science and Technology Beijing, Beijing, 100083, P.R. China

[3] Key Laboratory for Quantum Materials of Zhejiang Province, Department of Physics, School of Science, Westlake University, Hangzhou, Zhejiang 310024, China

[4] Shenyang National Laboratory for Materials Science, Institute of Metal Research, Chinese Academy of Sciences, Shenyang 110016, China

[5] Department of Electrophysics, National Yang Ming Chiao Tung University, Hsinchu, 30010 Taiwan

[6] National Synchrotron Radiation Research Center, 101 Hsin-Ann Road, Hsinchu 30076, Taiwan

[7] Materials Research Centre, Indian Institute of Science, Bangalore 560012, India

[†]These authors contributed equally to this work.

[*]Corresponding authors: E-mail: zuhuang@hit.edu.cn; liushi@westlake.edu.cn

## Abstract

The metastability of the polar phase in $HfO_2$, despite its excellent compatibility with the complementary metal-oxide-semiconductor process, remains a key obstacle for its industrial applications. Traditional stabilization approaches, such as doping, often induce crystal defects and impose constraints on the thickness of ferroelectric $HfO_2$ thin films. These limitations render the ferroelectric properties vulnerable to degradation, particularly due to phase transitions under operational conditions. Here, we demonstrate robust ferroelectricity in high-quality epitaxial $(HfO_2)_n/(ZrO_2)_n$ superlattices, which exhibit significantly enhanced ferroelectric stability across an extended thickness range. Optimized-period superlattices maintain stable ferroelectricity from up to 100 nm, excellent fatigue resistance exceeding $10^9$ switching cycles, and a low coercive field of ~0.85 MV/cm. First-principles calculations reveal that the kinetic energy barrier of phase transition and interfacial formation energy are crucial factors in suppressing the formation of non-polar phases. This work establishes a versatile platform for exploring high-performance fluorite-structured superlattices and advances the integration of $HfO_2$-based ferroelectrics into a broader range of applications.

# Introduction

The exploration of ferroelectric materials compatible with complementary metal-oxide-semiconductor (CMOS) technology is pivotal for advancing next-generation low-power nanoelectronic devices[1]. Since the discovery of ferroelectricity in fluorite HfO$_2$-based ultrathin films[2], exemplified by Hf$_{0.5}$Zr$_{0.5}$O$_2$ (HZO), these materials have become leading candidates for ferroelectric memory devices due to their exceptional scalability and compatibility with CMOS technology[3-5]. However, the energetically favorable phase in HfO$_2$ is the non-polar monoclinic phase (*m*-phase with $P2_1/c$ space group) rather than the desired metastable polar orthorhombic phase (*o*-phase with $Pca2_1$ space group). This disparity complicates the precise and controllable preparation of the polar phase, as polycrystalline and multiphase structures can easily compromise the reliability and reproducibility of ferroelectric devices. Consequently, achieving stable ferroelectric phase in HfO$_2$-based materials remains a challenging task[6-13], limiting the widespread industrial application of these materials.

Traditionally, stabilizing the metastable polar phase of HfO$_2$ relies heavily on the introduction of defects, such as oxygen vacancies[14,15]. However, optimal ferroelectric performance has only been reported in ultra-thin films (typically <10 nm thick)[10,16], with polarization diminishing significantly as film thickness increases and often vanishing beyond 20 nm[17,18]. Moreover, the type and concentration of defects can vary significantly with fabrication conditions, severely impacting overall ferroelectric performance[19,20]. For example, in practical applications, an excess of defects can lead to increased leakage current and promote the transition of metastable polar phase into non-polar phases, resulting in rapid degradation of ferroelectricity[21]. Consequently, these thin films often exhibit poor fatigue resistance during practical use. To overcome these challenges, innovative strategies for stabilizing the metastable polar phase are urgently required[8,22]. Drawing inspiration from successes observed in other oxide systems, constructing superlattices emerges as a promising approach. Superlattices enable precise control over critical variables such as lattice mismatch[23], local asymmetry[24], superlattice periodicity[25]. This level of control often leads to the manifestation of emergent phenomena,[26,27] such as unexpected ferroelectricity and multiferroicity[23,24], which are typically unattainable in chemically disordered solid-solution thin films.

While epitaxial fluorite-structured superlattices, particularly within the $HfO_2$-$ZrO_2$ system, hold immense potential for enhancing ferroelectricity and advancing the fundamental understanding of $HfO_2$-based ferroelectrics, their realization remains highly challenging[28-32]. Key obstacles include achieving precise layer control, ensuring interfacial stability, and maintaining phase purity. To date, there have been no reported studies on epitaxial $HfO_2$-based superlattices. Instead, previously reported multilayers and nanolaminates have predominantly exhibited polycrystalline structures[28,33-35], often accompanied by significant interfacial interdiffusion and disordered interfaces. These structural imperfections complicate the disentanglement of intrinsic interface-driven phenomena from extrinsic effects, hindering comprehensive insights into the role of interfaces in governing ferroelectric properties.

In this study, we successfully fabricated a series of high-quality, epitaxial ferroelectric superlattices of $(HfO_2)_n/(ZrO_2)_n$ ($n$ = 2, 3, 6, 11, and 16 unit cells) with well-defined interfaces, exhibiting a stable polar phase and enhanced ferroelectric properties. Systematic periodicity-dependent studies reveal that tuning the superlattice thickness effectively modulates the proportion of the metastable polar phase within the structure. Specifically, reducing the superlattice periodicity $n$ promotes polar phase formation in both $HfO_2$ and $ZrO_2$ layers, with an optimal periodicity $n$ of 3. As a result, these superlattices demonstrate robust ferroelectricity across a wide thickness range, significantly expanding the ferroelectric thickness window compared to conventional HZO solid-solution films, where ferroelectricity is typically confined to thicknesses below 10 nm[8,10,36,37,38]. At ultrathin scale (4 nm), the superlattices exhibit distinct macroscopic ferroelectricity. Remarkably, even at 100 nm, the superlattices exhibit robust ferroelectricity with a low coercive field of ~0.85 MV/cm. In addition, the superlattice also exhibits remarkable frequency and temperature stability. These results provide unparalleled flexibility in optimizing both thickness and ferroelectric performance in $HfO_2$-based films, offering opportunities for material design and device engineering. Furthermore, this work establishes a foundation for deeper investigations into the fundamental physics underlying ferroelectricity in hafnia-based systems.

## Results

**Epitaxial growth and structural characterizations of superlattices**

$(HfO_2)_n/(ZrO_2)_n$ ($n$ = 2, 3, 6, 11, and 16 unit cells) superlattices were epitaxially grown on (001)-oriented $La_{0.67}Sr_{0.33}MnO_3$ (LSMO)-buffered $SrTiO_3$ (STO) substrates at 600 °C (identified as the optimal condition for achieving ferroelectricity, Supplementary Fig. S1-S2) through pulsed laser deposition (Fig. 1a and Methods). The $\theta$–$2\theta$ X-ray diffraction (XRD) patterns of superlattices with various $n$ values are shown in Fig. 1b, and are compared with a random-alloy HZO film with an identical composition (~20 nm in total thickness). In addition to the expected (00$l$) STO and LSMO reflections, a prominent peak appears around 30° in all films, corresponding to the (111) reflection of the polar $o$-phase. Interestingly, despite having same total film thickness, the peaks near 28.2° and 34.5°, which correspond to the $m(\bar{1}11)$ and $m(200)$ crystallographic planes of the paraelectric $m$-phase, respectively, increase significantly when the superlattice periodicity $n > 6$. These XRD studies clearly demonstrate that smaller $n$ benefits the stabilization of the polar phase. The X-ray reflectivity (XRR) of the $n$ = 3 superlattice thin film (inset of Fig. 1b) shows clear Kiessig fringes, indicating the well-defined interfaces and high film quality.[39] XRR measurements of other superlattices also show sharp interfaces and well-defined periodicity (Supplementary Fig. S3). Additionally, atomic force microscopy topography image of the $n$ = 3 superlattice (Fig. 1c and Supplementary Fig. S4) reveals a smooth film surface with a root-mean-square roughness of only ~120 pm, further implying the high quality of the films. Furthermore, by calculating the area ratio of different diffraction peaks for each superlattice, we can qualitatively assess the trend of the $o$-phase variation with respect to the periodic thickness (Fig. 1d). This analysis clearly indicates that when $n$ < 6, the superlattice configuration exhibits a significantly higher proportion of the $o$-phase, which is noticeably higher than that of the HZO solid-solution film.

To gain a deeper understanding into the polar phase of the ~20-nm-thick superlattice with $n = 3$, XRD pole figure measurements were performed (Fig. 1e). The pole figure reveals 12 distinct diffraction spots around $\chi=71°$, with angular separations ($\Delta\varphi$) of 30°. This pattern arises from the presence of four crystallographic domains within the superlattice film, each rotated by 90° relative to the out-of-plane [111] direction. Each domain contributes three angular separations ($\Delta\varphi$) of 120° each, corresponding to diffraction spots arising from the $o(\bar{1}11)$, $o(1\bar{1}1)$, and $o(11\bar{1})$ crystal planes[10], confirming the presence of a well-defined polar $o$-phase. The pole figure results for the superlattice are consistent with those reported in $HfO_2$-based solid-solution thin films[10,16]. To further confirm the presence of polar $o$-phase in the superlattice, linearly polarized X-ray absorption spectroscopy (XAS) at the O $K$-edge was employed to probe the electronic structure of the films (Fig. 1f-g, and Supplementary Fig. S5). The O $K$-edge XAS investigates electronic transitions from the O $1s$ core level to the O $2p$ orbitals. Due to the strong hybridization between O $2p$ and Hf/Zr $5d$/$4d$ orbitals, the O $K$-edge XAS provides valuable information about the $d$-orbital electronic structure, predominantly within the energy range of 530 eV to 540 eV, as shown in Fig. 1f-g. Furthermore, the energy splitting within the $d$ orbitals manifests as a linear dichroism (XLD), providing insights into the anisotropic electronic structure of the films. Previous studies have demonstrated that such a polar phase of hafnium-based ferroelectrics can be characterized via the O $K$-edge XAS and XLD spectra[38]. The anisotropy in the XAS signals of the $n = 3$ superlattice, observed between out-of-plane and in-plane directions, along with the positive XLD signal at the $e_g$ manifold and negative signal at the $t_{2g}$ manifold, confirms the presence of polar phase (Fig. 1f)[28,38]. Interestingly, a reversed dichroism is observed in the ~20-nm-thick random-alloy HZO thin film (Fig. 1g), indicating a fundamentally different electronic structure in the solid-solution film compared to the superlattice. Previous studies have shown that the XAS spectra of $m$-phase $HfO_2$ exhibits a clear splitting of the $e_g$ states[40]. In contrast, the XAS spectra of pure $o$-phase $HfO_2$-based thin films display a single sharp $e_g$ orbital peak without any visible peak splitting[16]. The splitting of the $e_g$ orbitals in HZO solid solution film is evident in both $E // c$ and $E // ab$ orientations, along with an opposite XLD signal compared to the $n = 3$ superlattice in Fig. 1f, both providing the strong evidence for the existence of the $m$-phase in HZO (Fig. 1g)[40].

Atomic-scale cross-sectional transmission electron microscopy (TEM) characterizations offer a profound understanding of the microstructure of $HfO_2/ZrO_2$ superlattice thin films. Fig. 2a shows a high-angle annular dark-field scanning transmission electron microscopy (HAADF-STEM) image of the $n = 3$ superlattice, clearly showing distinct and well-defined interfaces between the $HfO_2$ and $ZrO_2$ layers. The inset fast Fourier transform (FFT) in the upper corner of Fig. 2a reveals the lattice orientations in reciprocal space, indicating the [111] out-of-plane direction of the superlattice. Notably, the superlattice reflections observed in the FFT (indicated by white arrows), extend clearly along the out-of-plane axis, signify a layer-by-layer growth mode with exceptional periodicity and highlighting the structural coherence of the superlattice. This is further corroborated by atomically resolved energy-dispersive X-ray spectroscopy (EDXS) mapping shown in Fig. 2b, which reveals distinct alternating $HfO_2$ and $ZrO_2$ layers along the out-of-plane direction. Only minimal mutual diffusion is observed at the interface, limited to a single atomic layer rather than forming an HZO solid solution. This indicates the precise compositional control during thin film growth. To gain deeper insights, the phase characteristics of both the $HfO_2$ and $ZrO_2$ layers were further analyzed. As shown in the enlarged views in Fig. 2c, both layers exhibit a distinct zig-zag arrangement feature of Hf or Zr atoms, being consistent with the structural characteristics of the $Pca2_1$ orthorhombic phase along the $[10\bar{1}]$ zone axis. This structural feature is maintained throughout the entire film along the thickness direction, providing additional evidence for the ferroelectric nature of the superlattice. This layered growth mode preserves the integrity of the polar phase throughout the film, affirming the robustness and stability of the superlattice architecture.

**Electrical characterizations of the superlattices**

Electrical characterizations of the superlattices provides crucial insights into their ferroelectric behavior. The polarization-electric field ($P$-$E$) hysteresis loops and corresponding switching current-electric field ($I$-$E$) curves for the ~20-nm-thick superlattices are presented in Fig. 3a and Fig. 3b, respectively. The results indicate that the $n = 3$ superlattice demonstrates the highest values of maximum polarization ($P_{max}$) of ~33 µC/cm² and remnant polarization ($P_r$) of ~15 µC/cm², accompanied by the most prominent switching current peak observed (Fig. 3b). As the superlattice periodicity $n$ increases, a noticeable degradation in the ferroelectric properties, characterized by a progressive reduction in $P_r$ and a diminishing switching current peak. The upper limit of periodicity for retaining ferroelectricity in the system is appears to be $n = 16$. At this periodicity, neither the $n = 16$ superlattice nor the corresponding HZO solid-solution film of equivalent thickness exhibit distinct polarization switching behavior, as indicated by the absence of pronounced current peaks in their $I$–$E$ characteristics. The $P_r$ of the $n = 2$ superlattice falls between those of the $n = 3$ superlattice and the HZO film. This is likely due to its exceedingly thin periodic thickness, which increases the likelihood of interfacial interdiffusion, thereby partially compromising its ferroelectric performance. Thus, in line with structural characterization findings, the $n = 3$ superlattice exhibits superior ferroelectric properties, attributed to its highest proportion of polar phase. Notably, the coercive field ($E_c$) of the superlattice thin films remains relatively constant across varying periodic thicknesses, ranging from 1.4 to 1.6 MV/cm. These values are lower than those reported for most HfO$_2$-based epitaxial films of similar thickness[10,16,18,41]. The slight shift observed in the $P$-$E$ hysteresis loops can be attributed to differences in the work functions between the top and bottom electrodes[42].

To further investigate the intrinsic ferroelectric behaviors of the films, $P$-$E$ curves were acquired through positive-up-negative-down (PUND) measurements (Supplementary Fig. S6). The $n = 3$ superlattice maintains a relatively high remnant polarization value even after non-ferroelectric contributions are excluded, highlighting its excellent ferroelectric performance. Additionally, permittivity-electric field ($\varepsilon$-$E$) characteristics further elucidated the variation in ferroelectric behavior across the superlattices (Fig. 3c). The $n = 2, 3, 6$, and 11 superlattices exhibit the characteristic 'butterfly' $\varepsilon$-$E$ hysteresis loops, affirming their ferroelectric nature. Meanwhile, the dielectric constant ($\varepsilon$) decreases with increasing $n$, which correlates with the observed reduction in the proportion of the polar $o$-phase. Since the $\varepsilon$ of the o-phase is higher than that of the $m$-phase[43], a lower $\varepsilon$ suggests a diminished presence of the polar phase. In contrast, the $\varepsilon$-$E$ curves for the HZO solid-solution film and the $n$ = 16 superlattice are nearly linear, indicating paraelectric behavior rather than ferroelectric behavior. For comparison, the electrical properties of 20-nm $HfO_2$ and $ZrO_2$ films were also measured, both of which exhibit typical paraelectric behavior, as shown in Supplementary Fig. S7. This confirms that, although achieving ferroelectricity in 20-nm-thick solid-solution (HZO) or single-component ($HfO_2$, $ZrO_2$) films is challenging, it can be readily realized through the construction of a superlattice structure. Furthermore, leakage current measurements reveal that an increased number of interfaces within the superlattice (i.e., reducing superlattice periodicity $n$) correlates with reduced leakage currents (Fig. 3d). This reduction arises from the interfaces, which effectively inhibit the migration of charged defects[44]. For example, the $n$ = 2 superlattice, with less distinct interfaces and notable interdiffusion at its boundaries, shows leakage current behavior similar to that of HZO. Comparative analysis of $2P_r$, $P_{max}$, and $\varepsilon$ (Fig. 3e) demonstrates that the ferroelectric performance of the superlattice improves as the periodic thickness decreases. Notably, the $n$ = 3 superlattice shows a remarkable enhancement in electrical characterizations over HZO solid solutions, with increases of 367%, 194%, and 71% in $2P_r$, $P_{max}$, and $\varepsilon$, respectively. These characteristics represent the most exceptional ferroelectric properties among $HfO_2$-based thin films studied.

In addition, we have conducted additional experiments to further validate the "super stable" ferroelectricity and scalability of superlattices. As shown in Fig. 3f, the 20 nm $n$ = 3 superlattice exhibits remarkable frequency stability within the range of 500 Hz to 50 kHz. In conventional ferroelectric oxide thin films, increasing frequency typically leads to a significant rise in coercive field[45,46]. However, frequency variation has virtually no impact on any physical quantities in $HfO_2$/$ZrO_2$ superlattice, including the coercive field, remnant polarization, or other parameters. More surprisingly, the superlattice demonstrates excellent temperature stability across a range of 77 K to 473 K. Compared to the retention of ferroelectricity at low temperatures, maintaining ferroelectricity under high-temperature conditions is evidently more relevant to the practical demands of ferroelectric memory devices. Previous works[18] have demonstrated the ability to maintain robust ferroelectricity only below room temperature, which clearly requires further improvement. By comparing the trends of $E_c$ and $2P_r$ with respect to frequency and temperature in the $n = 3$ superlattice (Fig. 3h), it is evident that frequency has almost no effect on $E_c$ and $2P_r$. Several studies have shown that ferroelectric switching in hafnia-based thin films is nucleation-limited, where domain nucleation, rather than domain wall motion, is the rate-limiting step[47-49]. This nucleation-limited switching (NLS) mechanism results in a higher coercive field, as domain nucleation is energetically more demanding than wall motion[50,51]. As a result, the field strength must exceed a critical threshold to overcome a substantial nucleation barrier, thereby intrinsically suppressing frequency dependence. This behavior contrasts sharply with that of conventional perovskite ferroelectrics, where domain wall motion is easily activated by comparatively weaker fields, making them more sensitive to frequency variations and external perturbations[52]. Those combined mechanisms offer a plausible explanation for the nearly frequency-independent behavior in Fig. 3f. Additionally, as the temperature increases, $2P_r$ exhibits only a slight increase, which can be attributed to the more significant movement of oxygen vacancies at higher temperatures[18]. Overall, these data highlight the unique advantages of the superlattice in maintaining ferroelectricity across varying frequencies and temperatures, and this "super-stable" characteristic will be advantageous for its wide application in electronic devices.

**Thickness variation ferroelectric properties of the superlattices**

Building on the foundation of periodic variations within the superlattice, we further investigated the effects of thickness evolution on overall ferroelectric properties. Superlattices with identical periodicity but different total thicknesses ($n = 3$ superlattices, with thicknesses ranging from 4 to 100 nm) exhibit significantly distinct ferroelectric properties (Fig.4). In the XRD patterns, the peak at $2\theta \approx 30°$ corresponds to the (111) plane diffraction of $o$-phase. Although the proportion of $m$-phase increases with increasing film thickness, peaks associated with the polar $o$-phase consistently remain predominant (Fig 4a). A noticeable expansion in the (111) lattice spacing ($d_{111}$) is evident in thinner films (Supplementary Fig. S8), indicating increased in-plane compressive strain, which plays a crucial role in stabilizing of polar phase[18]. In contrast, as thickness increases, the $d_{111}$ spacing remains relatively constant due to the strain relaxation in upper layers as the substrate constraint diminishes[10]. This relaxation promotes the formation of non-polar phases in thicker films. As a control, we also prepared random-alloy HZO solid-solution films with varying thicknesses, as shown in Supplementary Fig. S9a. A comparison of the XRD data reveals that the superlattices consistently maintain a significantly higher proportion of the polar phase than the HZO films of equivalent thickness, with the superlattices retaining a polar phase ratio exceeding 80% (Supplementary Fig. S9b). Notably, even the 100-nm-thick superlattice exhibits a higher proportion of polar phase than the 10-nm-thick HZO solid-solution film. This highlights the enhanced stabilization of the polar phase within the superlattices compared to HZO. Electrical characterization indicates that robust and reliable ferroelectricity still exists in the ultra-thin (4 nm) superlattice (Fig. 4b). This result demonstrates the high film quality and minimal defects of the 4 nm superlattice[16], which ensure the macroscopic ferroelectricity. In addition, at the 6 nm scale, the superlattice exhibits superior ferroelectricity relative to the HZO film, exhibiting larger $P_{max}$, comparable $P_r$, and a lower $E_c$ ((Supplementary Fig. S10a). Fig. 4c presents the endurance characteristics of the superlattice and HZO films following the wake-up process, with the inset showing the pulse sequences used for these endurance measurements. The superlattice maintained a stable $P_r$ value of ~ 10 μC/cm² (Supplementary Fig. S11a), with a variation of less than 15% over $10^9$ switching cycles. In contrast, the HZO film experienced dielectric breakdown after only $10^5$ cycles. Furthermore, a comparison of leakage currents before and after $10^5$ cycles (Supplementary Fig. S11b) reveals a significant increase in leakage current for the HZO film, whereas the superlattice film exhibited stable and nearly unchanged leakage current. This stability suggests that the superlattice interfaces effectively limit charged defect migration and charge injection, thereby contributing to the enhancement in cycling performance[53].

Beyond the enhanced ferroelectricity observed in ultrathin films, intriguing phenomena emerge in thicker films (10-100 nm). Remarkably, all superlattices, even at the film thickness up to 100 nm, exhibit well defined *P-E* hysteresis loops and prominent switching current peaks in the corresponding *I-E* curves (Fig. 4d). Moreover, the $E_c$ of these superlattices decreases with thickness increases, reaching a minimum value of ~0.85 MV/cm. In contrast, consistent with previous studies[18], as the thickness of HZO films increases to 20 nm, ferroelectricity nearly vanishes (Supplementary Fig. S12), thereby severely limiting the applicable thickness range of HZO. A comparison of remanent polarization in superlattices and HZO films across various thicknesses (Fig. 4e) highlights the superlattices' considerably higher polarization values and, more importantly, their broad adaptability. This marks the observation of ferroelectricity persisting at the 100 nm scale in $HfO_2/ZrO_2$ superlattice ($HfO_2$-based thin film)—a significant advancement surpassing the critical thickness threshold for maintaining ferroelectricity in HZO. Furthermore, as summarized in Fig. 4f[10,12,17,18,54-59], the coercive field of the superlattices in this study is substantially lower than those previously reported for epitaxial $HfO_2$-based thin film systems. This reduction in coercive field has important implications for lowering the energy consumption of memory devices, enhancing their overall efficiency and performance.

**The origin and stability of ferroelectricity in the superlattices**

Electrical characterizations indicate that individual $HfO_2$ and $ZrO_2$ films exhibit no significant ferroelectric behavior (Supplementary Fig. S7 and S10). However, the inverse size effect of fluorite ferroelectrics observed in previous studies suggests that ferroelectric behavior could emerge at ultrathin dimensions[38,60]. This raises an intriguing question: how does a stable polar phase exist in the $HfO_2/ZrO_2$ superlattices? First-principles calculations provide deeper insights into this phenomenon. While the energy barriers for phase transitions from the parent tetragonal phase (*t*-phase) to *o*- and *m*-phases in $ZrO_2$ are relatively low (Fig. 4g), the presence of the neighboring *o*-phase lattice of $HfO_2$ can significantly alter the kinetics. When constrained by the *o*-phase lattice of $HfO_2$, the kinetic energy barrier for the transition from the *t*-phase to the *m*- phase in ZrO$_2$ increases significantly (by approximately 273 meV/f.u.), while the barrier for the transition to the *o*-phase remains almost unchanged. This disparity in energy barriers favors the transformation of the high-temperature *t*-phase into the polar *o*-phase, thereby promoting the stabilization of a predominantly polar structure across both the HfO$_2$ and ZrO$_2$ layers within the superlattice.

With increasing film thickness, the proportion of the *m*-phase progressively rises, indicative of the inverse size effect commonly observed in HfO$_2$-based solid-solution films. This phenomenon is attributed to a thickness-dependent competition between the metastable *o*-phase and the thermodynamically stable *m*-phase. Previous theoretical investigations have proposed that substrate-induced strain plays a critical role in stabilizing the metastable *o*-phase in HfO$_2$, particularly in ultrathin films, by suppressing the formation of the energetically favored *m*-phase under specific conditions[22]. At lower thicknesses, the substrate exerts a significant influence on the structural phase of HfO$_2$, despite the *m*-phase being thermodynamically favorable in bulk. As film thickness increases, the clamping effect of the substrate on the HfO$_2$ film gradually diminishes. In a superlattice configuration, the growth of the *m*-phase on the *o*-phase lattice is further constrained by the *o-m* interfacial energy, which acts as an additional energetic barrier. These competing energy factors establish a critical thickness for the sustained growth of the *o*-phase within the superlattice. Beyond this critical thickness, the energetic advantage shifts, and the newly formed regions predominantly adopt the *m*-phase structure.

To further validate the proposed mechanism, we constructed the *o-m* interface structure for the (111) orientation (Supplementary Fig. S13). In a comparative HZO superlattice composed of alternating HZO layers of *o-* and *m-* phases, the *o-m* interface formation energy is approximately 64.6 meV/Å$^2$. By comparing the interface energy with the thermodynamic energy difference between the *m-* and *o*-phases, we determined a critical thickness of about 5.95 nm. It is noteworthy that, due to the elemental discontinuity, the *o-m* interface formation energy of the HfO$_2$-ZrO$_2$ superlattice (70.4 meV/Å$^2$) is higher than that of the HZO system. The critical thickness for the growth of ferroelectric ZrO$_2$ layer in the superlattice is determined to be 7.01 nm. This suggests that the elemental discontinuity inherent in the $HfO_2$-$ZrO_2$ superlattice structure is more effective in preventing the formation of the *m*-phase during film growth compared to HZO solid solution superlattice, which lacks a sharp elemental discontinuity. Consequently, $HfO_2$-$ZrO_2$ superlattices have the potential to maintain the dominance of the polar metastable phase even in thicker dimensions. Overall, the design of a superlattice structure elevates the formation energy of the *m*-phase structure at the interface of the *o*-phase substrate, thereby effectively mitigating the emergence of the *m*-phase as the film thickness increases

## Discussion

In summary, this study unveils a mechanism underlying the persistence of ferroelectricity in fluorite-structured $HfO_2$-based thin films. Through the strategic construction of $HfO_2$/$ZrO_2$ superlattices, composed of materials that are paraelectric in bulk, we achieve robust ferroelectricity over a wide thickness range (4 - 100 nm). The superlattice design notably reduces the coercive field to as low as ~0.85 MV/cm, presenting significant opportunities for lowering energy consumption in memory applications. Moreover, the superlattice also demonstrates excellent frequency and temperature stability. First-principles calculations reveal that the stabilization of the polar phase is governed by the elevated kinetic energy barriers for phase transitions and the interfacial formation energy, which collectively suppress the growth of non-polar phases. These findings provide a platform for precise control of ferroelectric properties in epitaxial $HfO_2$/$ZrO_2$ superlattices and highlight their potential in advancing both fundamental understanding and practical applications. The demonstrated adaptability of this system opens pathways for overcoming the polar phase instability and reducing coercive fields in $HfO_2$-based materials, pivotal steps toward the development of next-generation electronic devices and energy-efficient technologies.

## Methods

**$HfO_2$ based ferroelectric devices fabrication**.

HfO$_2$/ZrO$_2$ superlattices were grown on LSMO buffered-SrTiO$_3$ (001) single-crystal substrates by pulsed laser deposition (Arrayed Materials RP-B). The 10-nm-thick LSMO layer was deposited at a substrate temperature of 700 °C, employing a laser repetition rate of 3 Hz and a laser fluence of 0.8 J/cm$^2$. Subsequently, both HZO and HfO$_2$/ZrO$_2$ superlattices were deposited onto the LSMO layer at a temperature of 600 °C. The fabrication of superlattices was achieved by alternating growth sequences utilizing HfO$_2$ and ZrO$_2$ targets. For the ~20-nm-thick superlattices with $n$ = 2, 3, 6, 11, 16, the corresponding number of periods is 17, 10, 5, 3, and 2, respectively. And for the $n$ = 3 superlattices with thicknesses of 4, 6, 10, 20, 40, 60, and 100 nm, the corresponding number of periods is 2, 3, 5, 10, 22, 33, and 55, respectively. The laser fluences applied for the deposition of HZO, HfO$_2$, and ZrO$_2$ were all 1.3 J/cm$^2$. After deposition, the films were cooled down to room temperature at a cooling rate of 10 °C per minute, under a static oxygen pressure of 10$^4$ Pa. Circular top Pt electrode arrays with a diameter of 12.5 μm² and a thickness of 100 nm were fabricated for electrical measurements using photolithography and magnetron sputtering deposition (Array Materials RS-M).

**Structural characterizations.**

Detailed structural information was obtained using high-resolution X-ray diffraction (Rigaku SmartLab, 9 kW, Cu $K\alpha_1$ radiation) including $\theta$–$2\theta$ scans and pole figure. XRR measure was carried out at beamline BL02U2 of the Shanghai Synchrotron Radiation Facility (SSRF) (λ=1.239 Å). Atomic force microscopy (Asylum Research MFP-3D) was employed in tapping mode to monitor the surface morphology of the films.

**Soft X-ray spectroscopy measurements.**

Soft X-ray spectroscopy measurements were carried out at room temperature in the total electron yield (TEY) mode at the TLS11A and TPS45A beamlines of the National Synchrotron Radiation Research Center (NSRRC) in Taiwan. The soft X-ray beam was incident at an angle of 20° relative to the sample surface. The XLD signals were derived by subtracting the absorption spectra obtained with horizontal (*E//ab*) and vertical (*E//c'*, where *c'* represents an axis inclined 20° from the surface normal *c*) linearly polarized light.

**Scanning transmission electron microscopy characterizations.**

Scanning transmission electron microscopy characterizations, including HAADF-STEM imaging and EDXS mapping, were performed using an aberration-corrected FEI Titan Themis G2 microscope operated at 300 kV. This microscope is equipped with an X-FEG gun, monochromator, double aberration correctors, a GIF Quantum ER Energy Filter with DualEELS, and a Super-X EDXS system. For HAADF-STEM imaging, the convergent semi-angle and collection angle were set to 25 mrad and 48–200 mrad, respectively. Two-dimensional EDXS mapping was conducted based on the Super-X EDXS system, which provides high signal collection efficiency, enabling fast mapping with a high signal-to-noise ratio.

**Electrical properties measurements.**

Concerning electrical measurements were implemented via a metal-ferroelectric-metal (MFM) capacitance structure using the TF3000 analyzer (aixACCT). The MFM capacitances were subjected to electric fields, with the Pt electrode connected to the positive bias and LSMO grounded. Macroscopic electrical performances were including the *P-E* loops, *I-E* curves, $\varepsilon$–*E* characteristics, PUND measurement, leakage current and endurance et al. The macroscopic electrical performance of all samples was measured multiple times (> 4 times) at different capacitors to ensure the reliability and accuracy of the data.

**First-principles calculations.**

All first-principles density functional theory calculations were performed using Vienna Ab initio Simulation Package (VASP) with generalized gradient approximation of the Perdew-Burke-ErnZerhof (PBE) type. The $HfO_2$-$ZrO_2$ superlattice is modeled using a (111)-oriented supercell. *t*-phase of $ZrO_2$ was arranged at the top of o phase of $HfO_2$. We adjusted the number of Hf/Zr atomic layers to study the effect of different layer thickness. The plane-wave cutoff is set to 600 eV. We use a 2×2×1 Monkhorst-Pack k-point grid for structural optimizations.

# Data Availability

The source data in this study are provided in the Source Data file. Source data are provided with this paper.

## Code Availability

The code used to calculate the results shown in this work is available from the corresponding authors upon reasonable request.

## Acknowledgments

This work was supported by National Key R&D Program of China (Grant No. 2021YFA1202100), National Natural Science Foundation of China (Grant Nos. 92477129, 52372105, 12361141821), Guangdong Basic and Applied Basic Research Foundation (Grant No. 2024B1515120010), and Shenzhen Science and Technology Program (Grant No. KQTD20200820113045083). Z.H.C. has been supported by State Key Laboratory of Precision Welding & Joining of Materials and Structures (Grant No. 24-Z-13) and "the Fundamental Research Funds for the Central Universities" (Grant No. 2024FRFK03012). The authors thank the staff from Shanghai Synchrotron Radiation Facility (SSRF) at BL02U2.

## Author Contributions

Z.H.C. and J.X.L. conceived and designed the experiments. Z.H.C. supervised the study. J.X.L. fabricated the films and performed electrical measurements and data analysis with assistance from Y.Y.S., C.Z., S.Z.H. and Y.J.L.. J.X.L. conducted XRD measurements with help from Y.Y.S. S.Q.D., Y.L.T., and K.F.W. contributed to the STEM characterizations. L.Y.M., J.Y.Y. and S.L. carried out theoretical calculations. Y.-C.K. and C.-Y.K. performed the soft spectroscopy measurements. J.X.L., S.Q.D., L.Y.M., S.D., S.L. and Z.H.C. wrote the manuscript. All authors discussed the results and commented the manuscript.

## Competing Interests

The authors declare no competing interests.

## Additional information

**Supplementary information**
Supplementary information Fig. S1-S13

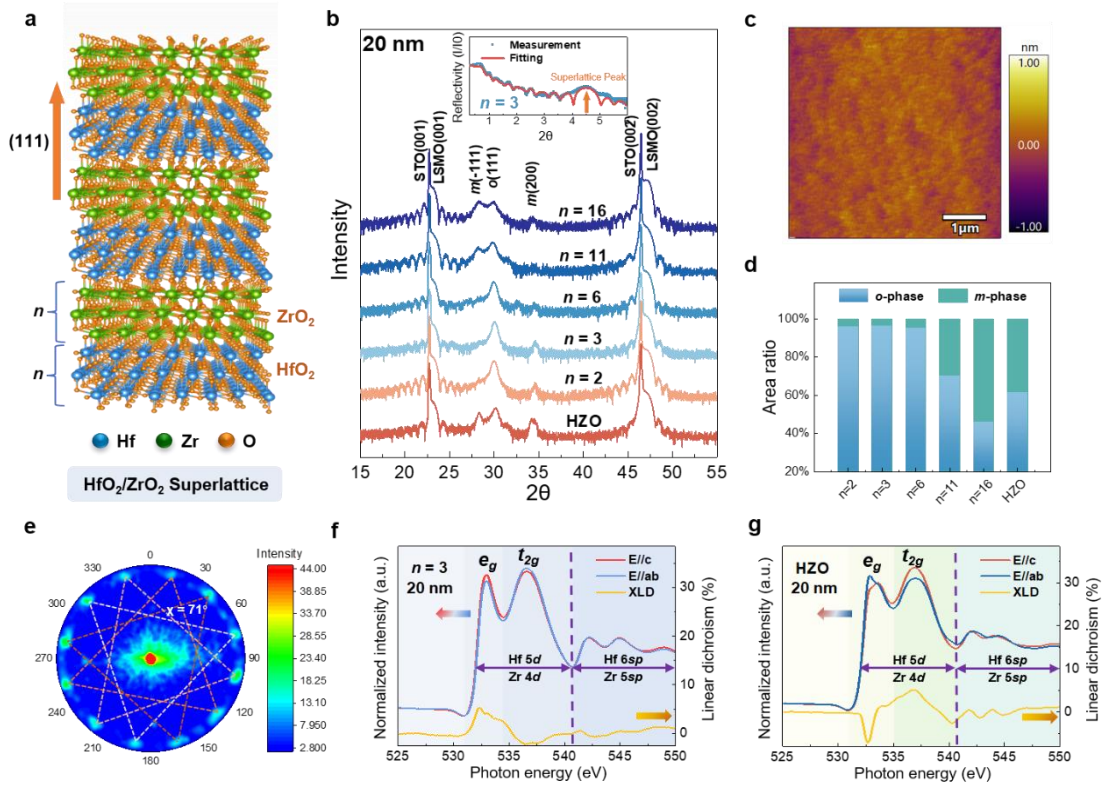

**Figure 1. Structural characterizations of ~20-nm-thick epitaxial $HfO_2/ZrO_2$ superlattices. a,** Schematic illustration of the lattice structure of $HfO_2/ZrO_2$ superlattices deposited on STO/LSMO (Methods). The same composition ratio was used to create a direct comparison between the ordered $[HfO_2]_n/[ZrO_2]_n$ superlattices and the disordered $Hf_{0.5}Zr_{0.5}O_2$ solid-solution films. **b,** XRD patterns of 20 nm superlattices with varying periodicity. The inset display the XRR pattern of the $n = 3$ superlattice, and clear superlattice reflection can be observed. **c,** Representative AFM topographic image of the $n = 3$ superlattice. **d,** The peak area ratio of different phases extracted from XRD patterns. The area ratios of the $o$-phase and $m$-phase peaks were calculated by $I_{o(111)}/(I_{m(-111)} + I_{m(200)} + I_{o(111)})$ and $(I_{m(-111)} + I_{m(200)})/(I_{m(-111)} + I_{m(200)} + I_{o(111)})$, respectively. **e,** Pole figure of the $n = 3$ superlattice, showing its crystallographic orientation. **f-g,** XAS and XLD at the O K-edge for the (**f**) $n = 3$ superlattice and (**g**) corresponding the 20 nm HZO solid solution film. The shaded background areas, differentiated by distinct colors, represent discrete crystal field configurations within the materials.

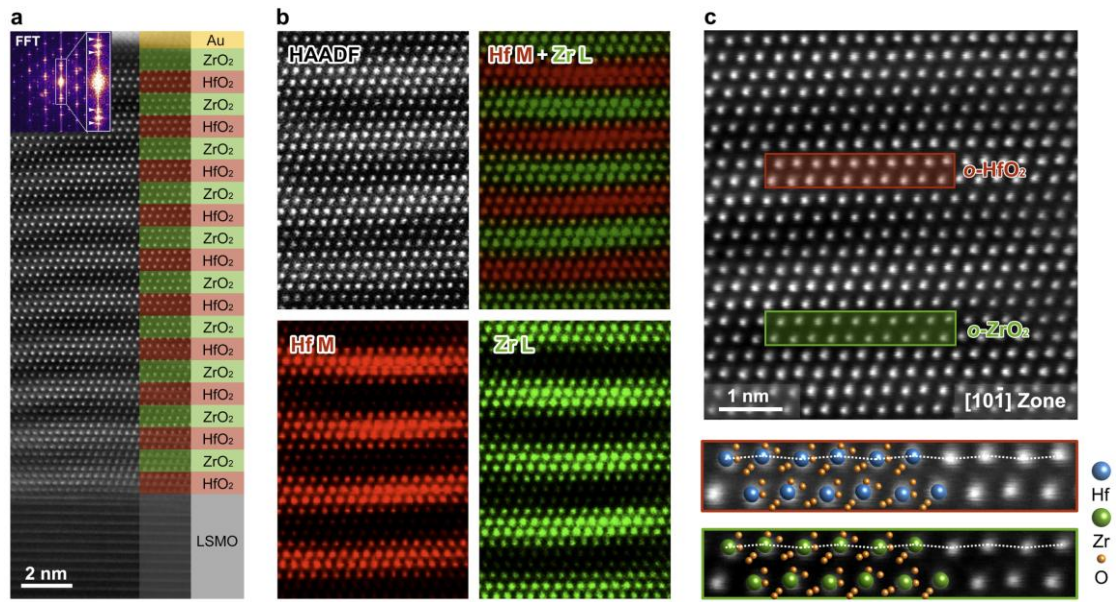

**Figure 2. Atomic-resolution STEM studies of the ~20-nm-thick $HfO_2/ZrO_2$ superlattice with $n = 3$. a**, Atomically resolved HAADF-STEM image of the superlattice film. Inset shows the corresponding fast Fourier transform of the image, highlighting superstructure spots indicated by white arrows. **b,** Atomic-scale EDXS mapping of Hf M and Zr L- signals, revealing a chemically well-defined superlattice structure. **c,** Enlarged view of the $HfO_2$ and $ZrO_2$ layers, both manifesting an evident zig-zag arrangement of Hf or Zr atomic layers, characteristic of the *o*-phase.

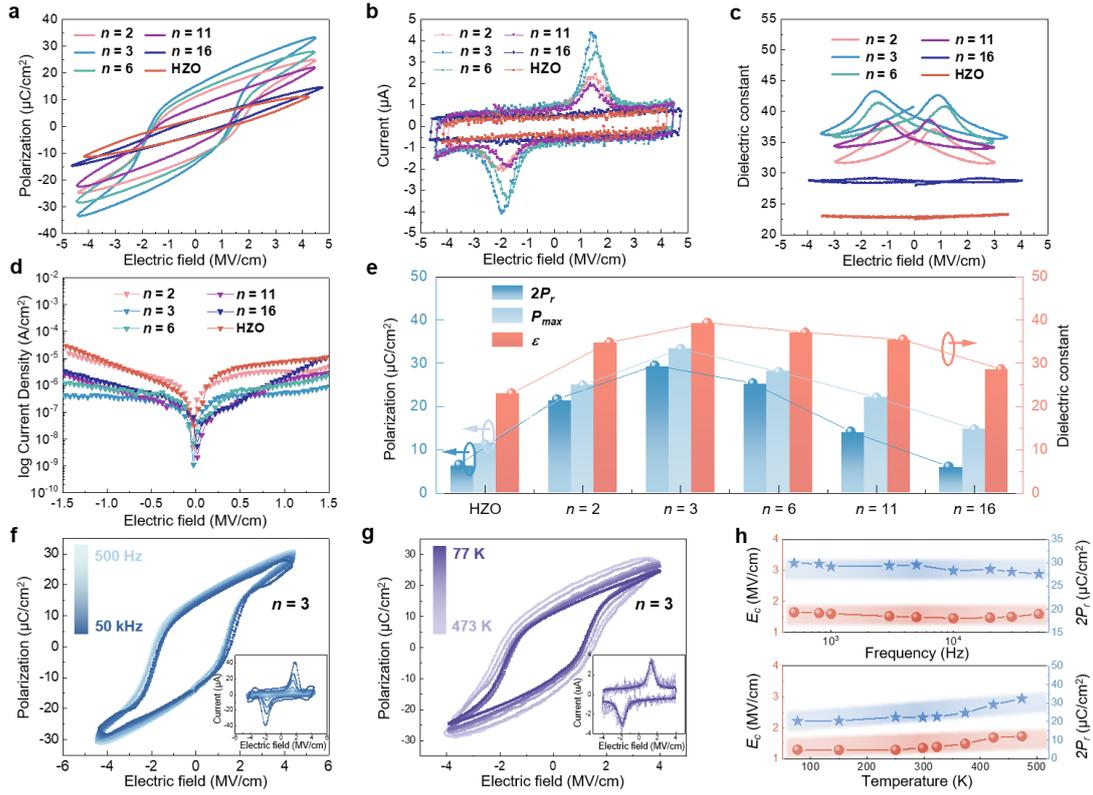

**Figure 3. Electrical properties of ~20-nm-thick HfO$_2$/ZrO$_2$ superlattices as a function of periodicity. a-b,** (**a**) *P-E* hysteresis loops and (**b**) corresponding *I-E* curves under an electric field of frequency 5 kHz. **c-d,** (**c**) Dielectric constant and (**d**) leakage current as a function of electric field. **e,** Comparison of 2$P_r$, $P_{max}$ and $\varepsilon$ of the 20 nm superlattice and HZO solid-solution films. **f-g,** *P-E* hysteresis loops and corresponding *I-E* curves (inset) of *n* = 3 superlattice at (f) different frequencies (500 Hz - 50 kHz) and (g) different temperatures (77 K - 473 K). **h,** The trend of the variation of $E_c$ and 2$P_r$ with frequency and temperature in *n* = 3 superlattice.

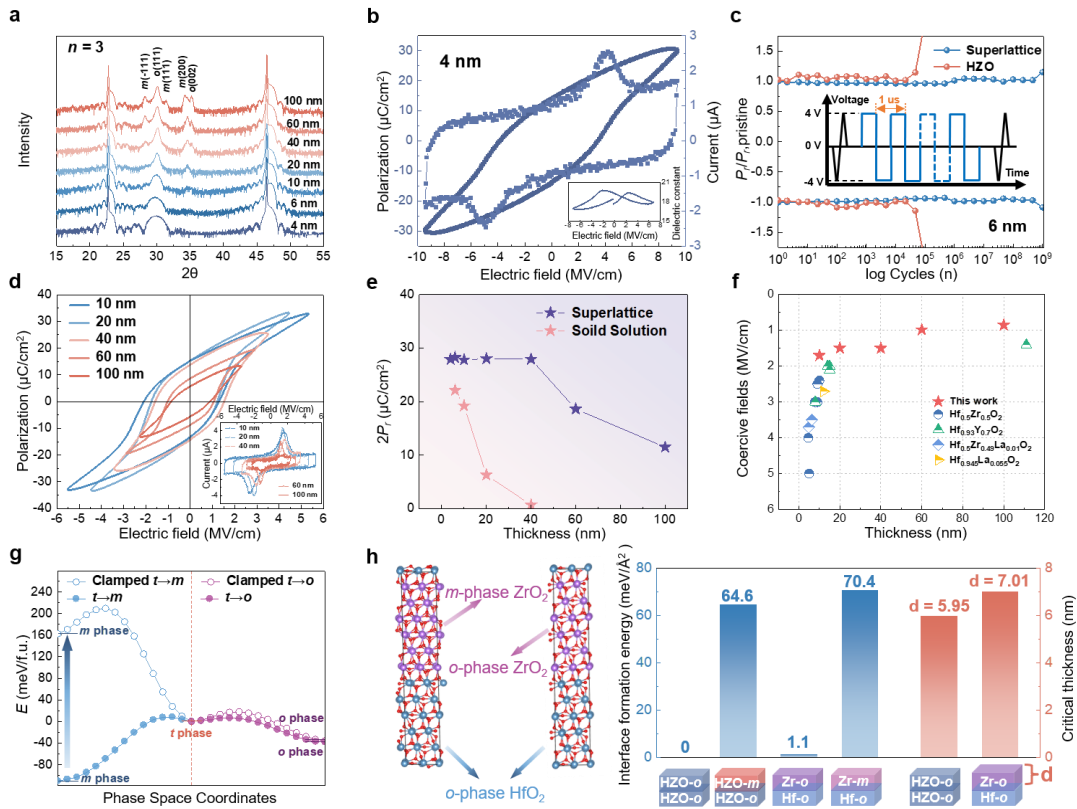

**Figure 4. Robust ferroelectricity in HfO$_2$/ZrO$_2$ superlattices and its microscopic origin. a,** XRD patterns of the *n* = 3 superlattices with different total film thickness (4 -100 nm). **b**, *P-E* hysteresis loop, corresponding *I-E* curve and dielectric constant-electric field curve (inset) of 4 nm superlattice, which highlights the ferroelectricity of the superlattices at ultra-thin scales. **c,** Endurance performances of 6 nm HZO and superlattice. **d,** *P-E* hysteresis loops and corresponding *I-E* curves (inset) of 10 nm, 20 nm, 40 nm, 60 nm and 100 nm superlattices. **e,** Remnant polarization of superlattices and HZO films at different thickness. **f,** Comparison of minimum coercive fields of our superlattices with representative reported epitaxial HfO$_2$-based ferroelectric materials. **g,** Phase transition kinetics energy barrier of zirconia. **h,** Schematic of the atomic interfaces model (left). The interfacial forming energy of different systems and critical thickness of polar phase (right).